\documentclass[aps,prx,twocolumn,natbib,showpacs,floatfix]{revtex4-1}

\usepackage{graphicx}
\usepackage{amsmath,amssymb,bbold,bm,color}
\usepackage{float}
\usepackage{epstopdf}
\usepackage{hyperref}
\usepackage{color}
\usepackage{enumerate}
\usepackage{wasysym}
\usepackage{url}
\usepackage{dsfont}
\usepackage{braket}
\hypersetup{colorlinks=true,linkcolor=blue,citecolor=blue,urlcolor=blue}
\graphicspath{{figures/}{../figures/}}

\newcommand{\bk}{{\bm k}}

\newcommand{\bq}{{\bm q}}

\newcommand{\br}{{\bm r}}

\newcommand{\cF}{{\cal F}}

\newcommand{\cT}{{\cal T}}

\newcommand{\cH}{{\cal H}}

\newcommand{\cN}{{\cal N}}
\newcommand{\cV}{{\cal V}}

\newcommand{\bee}{\begin{equation}}
\newcommand{\ee}{\end{equation}}
\renewcommand{\bra}[1]{\langle #1 |}
\renewcommand{\ket}[1]{| #1 \rangle}

\begin{document}

\title{Josephson effects in twisted cuprate bilayers}

\author{Tarun Tummuru}
\author{Stephan Plugge}
\author{Marcel Franz}

\affiliation{Department of Physics and Astronomy, and Quantum Matter
  Institute, University of British Columbia, Vancouver, BC, Canada V6T 1Z1}

\begin{abstract} 
  Twisted bilayers of high-$T_c$ cuprate superconductors have been argued to form topological phases with spontaneously broken time reversal symmetry $\cT$ for certain twist angles. With the goal of helping to identify unambiguous signatures of these topological phases in transport experiments, we theoretically investigate a suite of Josephson phenomena between twisted layers.
  We find an unusual non-monotonic temperature dependence of the critical current at intermediate twist angles which we attribute to the unconventional sign structure of the $d$-wave order parameter. The onset of the $\cT$-broken phase near $45^\circ$ twist is marked by a crossover from the conventional $2\pi$-periodic Josephson relation $J(\varphi)\simeq J_c\sin{\varphi}$ to a $\pi$-periodic function as the single-pair tunneling becomes dominated by a second order process that involves two Cooper pairs. Despite this fundamental change, the critical current remains a smooth function of the twist angle $\theta$ and temperature $T$ implying that a measurement of $J_c$ alone will not be a litmus test for the $\cT$-broken phase.
  To obtain clear signatures of the $\cT$-broken phase one must  measure $J_c$ in the presence of an applied magnetic field or radio-frequency drive, where the resulting Fraunhofer patterns and Shapiro steps are altered in a characteristic manner. We discuss these results in light of recent experiments on twisted bilayers of the high-$T_c$ cuprate superconductor Bi$_2$Sr$_2$CaCu$_2$O$_{8+\delta}$.
\end{abstract}

\date{\today}
\maketitle

\section{Introduction}

Dissipationless current driven by a phase difference between
superconductors  -- the Josephson effect \cite{Josephson1962} -- constitutes a
quintessential manifestation of the coherent quantum state on a
macroscopic scale. Measurements of the critical current $J_c$ as a function of
temperature $T$, magnetic field $B$ and applied electromagnetic
radiation provide important insights into the phenomenology and
microscopic mechanisms of a great variety of superconductors ranging
from conventional metals to cuprate and iron-based unconventional
superconductors. In this paper we discuss various aspects of the
Josephson effect between two samples of high-$T_c$ cuprates stacked
with a relative twist angle $\theta$. It is well known that cuprates
are $d$-wave superconductors \cite{Tsuei2000}, implying a strongly anisotropic order
parameter. One thus expects significant dependence of the Josephson
current on the twist angle, which was indeed noted in the classic works
on this topic~\cite{Klemm1998,Klemm2000,Klemm2001}. For reasons that remain not well understood, this expectation was not borne out in early experimental works \cite{LI19971495, Li1999, Takano_2002}, which showed twist-independent critical currents. Very recent experimental work on carefully assembled thin Bi$_2$Sr$_2$CaCu$_2$O$_{8+\delta}$ (Bi2212) flakes \cite{Zhao2021} finally reported the critical current anisotropy indicative of $d$-wave order parameter (see however Ref.\ \cite{Xue2021} for a recent contrary result). 

\begin{figure}[b]
  \includegraphics[width=8cm]{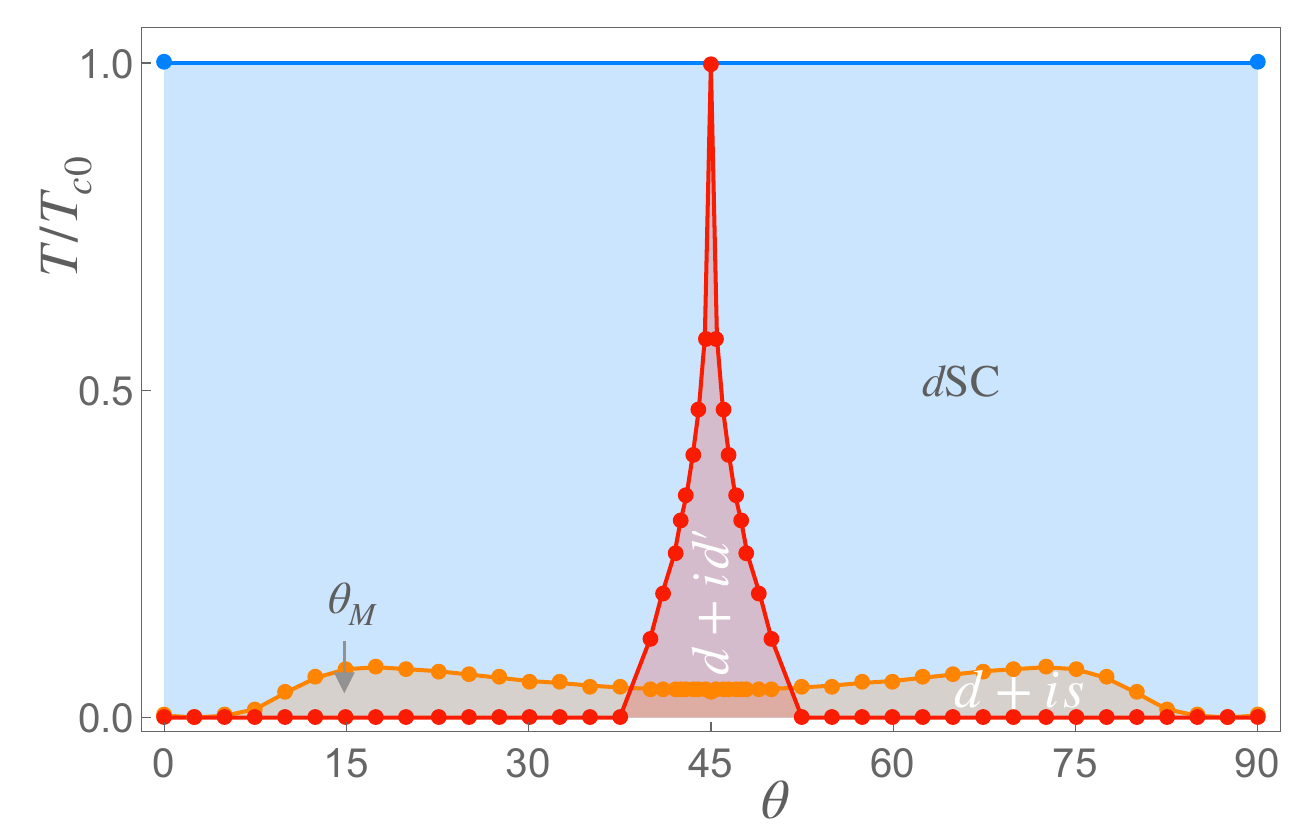}
  \caption{Typical phase diagram of a twisted cuprate bilayer
    calculated from the microscopic model discussed in Sec.\ \ref{sec:model}. It is
  assumed that $d+is$ order is nucleated near the magic angle
  $\theta_M$ with the $s$ component driven by a weak on-site
  attractive interaction. Model parameters are $\epsilon_c=60$meV,
  $g=10.5$meV, $N_F\cV=0.12$ and $N_F\cV_s=0.043$. }
  \label{fig1}
\end{figure}

The experimental finding \cite{Yuanbo2019} that high-$T_c$ cuprate Bi2212 superconducts up to $T_c\simeq 90$K when exfoliated down to a single monolayer inspired renewed interest in twisted Bi2212 {\em bilayers}. 
Theoretical studies predict the emergence of novel spontaneously $\cT$-broken phases
with full excitation gap and nontrivial topology near the 45$^{\rm o}$ twist \cite{Can2021} as well as in the
vicinity of the `magic angle' $\theta_M$ \cite{volkov2020magic} that depends on model parameters but is generally in the range of few degrees. The physics of the latter
bears some resemblance to the magic angle in twisted bilayer graphene
\cite{Bistritzer2011,Cao2018a,Cao2018b,Yankowitz2019,Sharpe2019,Efetov2019}
except that flat bands are replaced by a quadratic band crossing that
arises from the collision of two Dirac nodes. The $\cT$-breaking near
45$^{\rm o}$ arises from the coherent tunneling of pairs of Cooper
pairs when ordinary single-pair tunneling is suppressed by the
$d$-wave symmetry of the pair wavefunction. A typical phase diagram,
predicted on the basis of continuum Bogoliubov-de Gennes (BdG) theory
discussed in detail below, is shown in Fig.\ \ref{fig1}. It
illustrates the central theoretical finding that at the 45$^{\rm o}$ twist the
topological $\cT$-broken $d+id'$ phase has a potential to persist up
to high temperature, approaching the native $T_c$ of the cuprate
material. 

The goal of this paper is to elucidate how these predicted $\cT$-broken phases are 
manifested in the Josephson effect and, specifically, what type of
measurement can deliver solid evidence for their existence. We also
address the fate of the $\cT$-broken phases in twisted structures
composed of thicker flakes which might be easier to assemble in the
laboratory. Our main finding in this regard is that the basic
phenomenology of $\cT$-breaking remains in place although the
topological gap is strongly suppressed as the number of
monolayers comprising the flake increases.

Much of the phenomenology underlying the Josephson effect can be understood
on the basis of the Ginzburg-Landau (GL) theory where each layer is described by a complex scalar order parameter. For two weakly coupled
monolayers, the corresponding GL free energy takes the form
\begin{eqnarray}
 \cF[\psi_1,\psi_2]&=&\cF_0[\psi_1]+\cF_0[\psi_2]+A |\psi_1|^2
                      |\psi_2|^2 \label{e1}\\
  &+&B(\psi_1\psi_2^\ast +{\rm c.c.})+C(\psi_1^2\psi_2^{\ast 2} +{\rm c.c.}),\nonumber
\end{eqnarray}
where $\psi_{a}$ represent the complex order parameters in layers $a=1,2$ and
\begin{equation}\label{e2}
\cF_0[\psi]=\alpha|\psi|^2+{1\over 2}\beta|\psi|^4
\end{equation}
is the free energy of a monolayer. Terms on the second line of Eq.\ \eqref{e1} will be seen to underlie the
interlayer Josephson effect. Physically, the $B$ and $C$ terms
represent, respectively, coherent tunneling of single and double Cooper
pairs between the layers. 

For two identical layers, the order parameters can only differ by a phase which allows us to write
\begin{equation}\label{e3}
\psi_1=\psi, \ \ \ \psi_2=\psi e^{i\varphi},
\end{equation}
where we take $\psi$ to be real and positive. $d$-wave
symmetry additionally implies that GL parameter $B$ must change sign when
the twist is increased by $\pi/2$. We henceforth assume the
simplest angle dependence consistent with this condition, $B=-B_0\cos(2\theta)$,
where $B_0>0$ is taken to ensure that at zero twist layers are
in-phase. It is also generally true that $C>0$. With these ingredients we may write the corresponding Josephson free energy 
\begin{equation}\label{e4}
\cF(\varphi)={\cal E}_0-{\hbar\over 2e}\left[J_{c1}\cos{\varphi}-{1\over 2} J_{c2}\cos({2\varphi)}\right],
\end{equation}
where $ J_{c1}=(4eB_0/\hbar)\psi^2\cos(2\theta)$, $J_{c2}=(2eC/\hbar)\psi^4$ and ${\cal E}_0$ collects terms that are
independent of $\varphi$. The competition between the $\cos\varphi$
and $\cos 2\varphi$ terms in Eq.\ \eqref{e4} underlies the emergence
of the spontaneously $\cT$-broken phase near the 45$^{\rm o}$ twist.
When $\theta$ is close to zero the conventional Josephson tunneling term $J_{c1}$ dominates and the free energy minimum occurs at $\varphi=0$. 
Increasing the twist, however, decreases $J_{c1}\sim\cos(2\theta)$. Eventually, for twist angle approaching $45^\circ$, the $J_{c2}$ term begins to dominate, and
$\cF(\varphi)$ develops two distinct minima at $\pm\varphi_{\rm min}$ signalling
the $\cT$-broken phase. 

The other $\cT$-broken phase that occurs near the magic
angle $\theta_M$ depends on the structure of the low-energy quasiparticle excitations and interaction physics. As such it 
cannot be understood based on the simple GL theory formulated above and we will review its origin below.  

The equilibrium current between the layers follows from the Josephson
relation
\begin{equation}\label{e5}
  J(\varphi)=(2e/\hbar)d\cF/d\varphi,
\end{equation}
which yields a simple but all-important current-phase relation
\begin{equation}\label{e6}
J(\varphi)=J_{c1}\sin{\varphi}- J_{c2}\sin({2\varphi)}
\end{equation}
that will form the basis for much of our analysis. It
is to be noted that coefficients $J_{c1,2}$ depend on both the twist
angle $\theta$ and temperature $T$. Within the basic GL theory this
dependence follows from expressions below Eq.\ \eqref{e4} together
with $\psi(T)=\psi_0\sqrt{1-T/T_c}$, which holds close to $T_c$. More
accurate dependencies valid for all temperatures can be obtained from
a microscopic model that will be discussed in the next Section. We note that there have been studies that have looked at the general consequences of the presence of a second harmonic in the current-phase relation (cf. Ref.~\cite{Goldobin_2007}). Since our parameters are informed by the microscopic model, our results will be more directly relevant to twisted cuprates. 

The physical observable that is most straightforward to measure experimentally is
the critical current
\begin{equation}\label{e7}
  J_c=\max_\varphi[ J(\varphi)].
\end{equation}
We find that $T$ and $\theta$ dependence of $J_c$ show some interesting features
in the low-$T$ regime but, perhaps surprisingly, do not contain any clear signatures
of the $\cT$-broken phases; $J_c(T,\theta)$ is a smooth function of
its arguments across the transition to the $\cT$-broken phase. The transition is signalled by the vanishing phase stiffness, 
defined as 
\begin{equation}\label{e8}
 \rho_s={d J(\varphi)\over d\varphi}\biggr|_{\varphi=0}.
\end{equation}
Mathematically, the condition $\rho_s=0$ marks the point at which the free energy minimum at $\varphi=0$ becomes a local maximum.  
Unfortunately, $\rho_s$ is not easily measurable. For the
conventional sinusoidal current-phase relationship (i.e.\ when
$J_{c2}=0$), it is easy to see that $\rho_s$ and $J_c$  coincide.   
We will highlight departures from the $J_c=\rho_s$ equality to quantify
deviations from the conventional Josephson behavior in twisted cuprate
bilayers.

Writing $J_{c1}(\theta)=J_{c1}(0)\cos(2\theta)$, Eq.\ \eqref{e8} can be used to determine the critical twist angle $\theta_c$ beyond which $\cT$ breaking occurs, 
\begin{equation}\label{e88}
\theta_c={1\over 2}\arccos{\left({}2J_{c2}\over J_{c1}(0)\right)}. 
\end{equation}
We recall that, within GL theory, $J_{c2}$ is a constant independent of $\theta$.

When magnetic field $B$ is applied parallel to the plane of the
junction the phase difference between the layers becomes space
dependent and gives rise to the well-known Fraunhofer oscillations with the
critical current given by
\begin{equation}\label{e9}
 J_c(\Phi)=J_c(0)\left|{\sin(\cN\pi\Phi/\Phi_0)\over \cN \pi\Phi/\Phi_0}\right|.
\end{equation}
Here $\Phi=BS$ is the flux through the effective junction area $S$
and $\Phi_0=hc/2e$ is the superconducting flux quantum. For an ordinary
Josephson relation we have $\cN=1$. On the other hand, when the current-phase relation \eqref{e4} is
dominated by the $\sin(2\varphi)$ term we expect $ J_c(\Phi)$ to
follow Eq.\ \eqref{e9} except with $\cN=2$. When both terms are present in the current-phase relation a more detailed analysis,
given in Sec.\ \ref{sec:fraunhofer},
predicts a crossover from the conventional $\cN=1$ Fraunhofer
pattern at small twist angles to the $\cN=2$ behavior as $\theta$
approaches $45^{\rm o}$. We conclude that the field dependence of the
critical current can be tested to probe for spontaneous $\cT$ breaking
in twisted cuprate bilayers. Similar results are obtained for Shapiro steps in Sec.\ \ref{sec:shapiro}, where our theory
predicts that fractional steps appear in the current-voltage characteristics 
when a $\cT$ broken state is subjected to an external bias in the form of radio-frequency radiation.


\section{Microscopic model}
\label{sec:model}
 
In order to capture the essential physics of the twisted bilayer system we work with a continuum microscopic model of coupled $d$-wave superconducting monolayers \cite{Can2021} which we check against the more accurate lattice model. Within the continuum model it is straightforward to obtain the current-phase relation for any temperature $T$ and twist angle $\theta$ which we in
turn use to extract coefficients $J_{c1,2}$ that enter the phenomenological Eq.\ \eqref{e6}. In the subsequent Sections, Eq.\ \eqref{e6} is then employed to make detailed predictions for Fraunhofer oscillations and Shapiro steps that can be experimentally probed to reveal the presence of spontaneous $\cT$ breaking in the system.

The model is defined by the second-quantized Hamiltonian 
\begin{equation}\label{h3}
\cH=\sum_\bk\Psi_\bk^\dag H_\bk \Psi_\bk+E_0
\end{equation}
where $\Psi_\bk=(c_{\bk\uparrow  1}, c^\dag_{-\bk\downarrow 1}, c_{\bk\uparrow 2},c^\dag_{-\bk\downarrow 2})^{T}$ represents a four-component Nambu spinor  and
$c^\dag_{\bk\sigma a}$ creates an electron with spin $\sigma$ in layer
$a$ of the twisted bilayer. The BdG Hamiltonian is a $4\times 4$ matrix
\begin{equation}\label{h4}
  H_\bk=
 \begin{pmatrix}
   \xi_\bk & \Delta_{\bk 1} & g & 0 \\
   \Delta_{\bk 1} ^\ast & -\xi_\bk & 0 & -g \\
   g & 0 &   \xi_\bk & \Delta_{\bk 2} \\
   0 & -g &    \Delta_{\bk 2} ^\ast & -\xi_\bk
  \end{pmatrix}
\end{equation}
with normal-state dispersion $\xi_\bk=\hbar^2 k^2/2m-\mu$, interlayer
coupling $g$, and $d$-wave pairing amplitudes
\begin{eqnarray}
\Delta_{\bk 1}&=& \Delta e^{i\varphi/2}\cos(2\alpha_\bk-\theta), \label{h5}\\
\Delta_{\bk 2}&=& \Delta e^{-i\varphi/2}\cos(2\alpha_\bk+\theta). \label{h6}       
\end{eqnarray}
Here $\Delta$ is taken as real positive, $\alpha_\bk$ denotes the
polar angle of vector $\bk$, and $\pm\theta$ terms
encode the twist between the layers. Finally, 
\begin{equation}\label{h7}
E_0=\sum_\bk 2\xi_\bk -{1\over\cV}\sum_{\bk  a}|\Delta_{\bk a}|^2, 
\end{equation}
where the last term results from the standard mean-field decoupling of
the pairing interaction in the $d$-wave channel with strength
$\cV$. The corresponding free energy is given by
\begin{equation}\label{h8}
 \cF_{\rm BdG}=E_0-2 k_BT\sum_{\bk \alpha}\ln\left[2\cosh{(E_{\bk \alpha}/2 k_BT)}\right].
\end{equation}
where the sum extends over all positive energy eigenvalues $E_{\bk \alpha}$ of $H_\bk$.

\begin{figure}[t]
  \includegraphics[width=8cm]{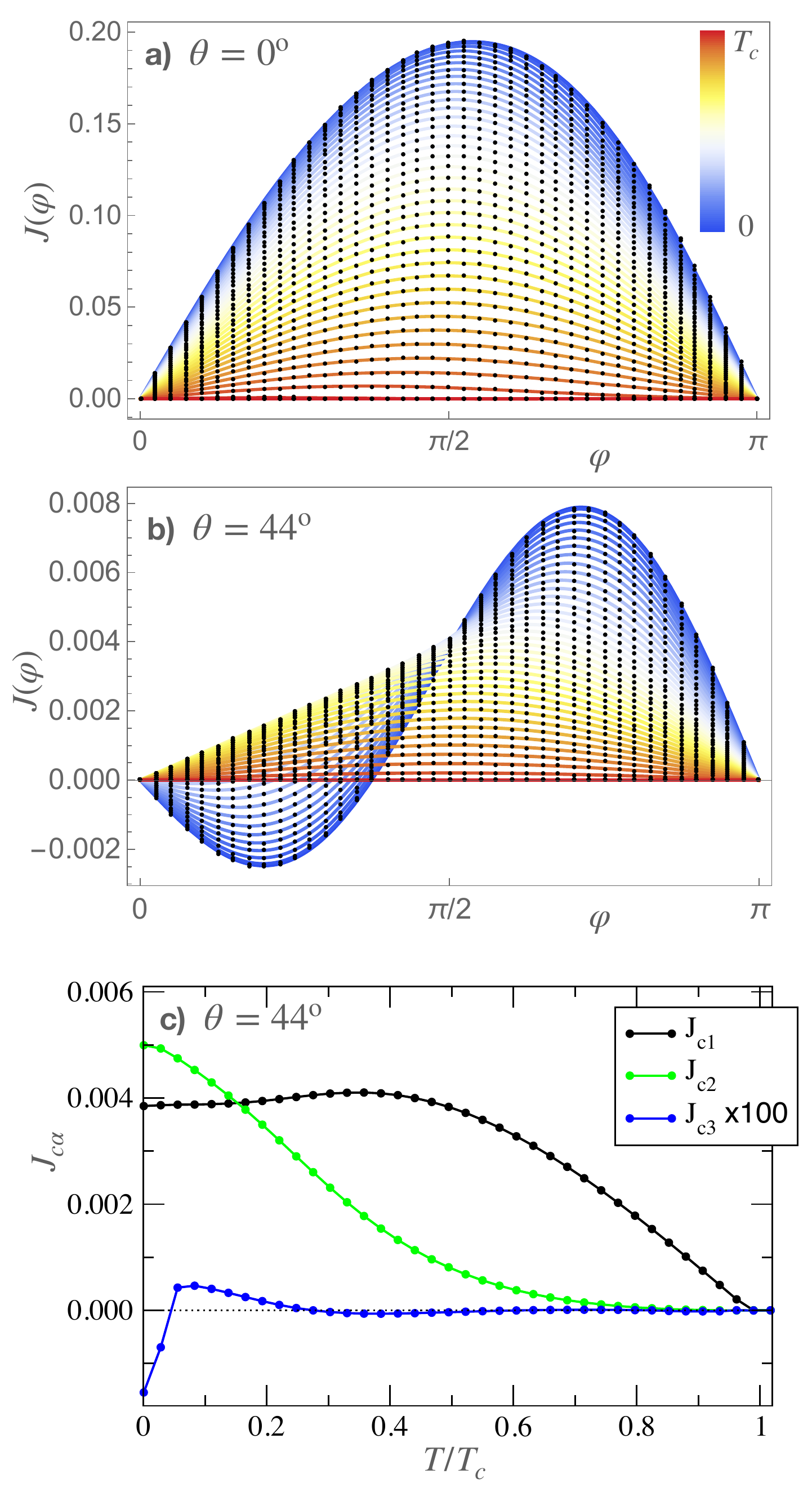}
  \caption{(a,b) Typical current-phase relations obtained from the continuum model. Data points represented as black dots are obtained by evaluating Eq.\ \eqref{h13} at fixed twist angle $\theta$ for a range of temperatures between zero and just above $T_c$. The color-coded solid lines represent the best fit to the relation Eq.\ \eqref{e6} amended by a third order term $-J_{c3}\sin(3\varphi)$ to achieve a close fit. Model parameters are $\epsilon_c=60$meV,
  $g=7$meV, $N_F\cV=0.12$. Panel (c) shows the temperature dependence of the best-fit coefficients $J_{c\alpha}$ $(\alpha=1,2,3)$ for $\theta=44^{\rm o}$.}
  \label{fig3}
\end{figure} 

\begin{figure*}[t]
  \includegraphics[width=17cm]{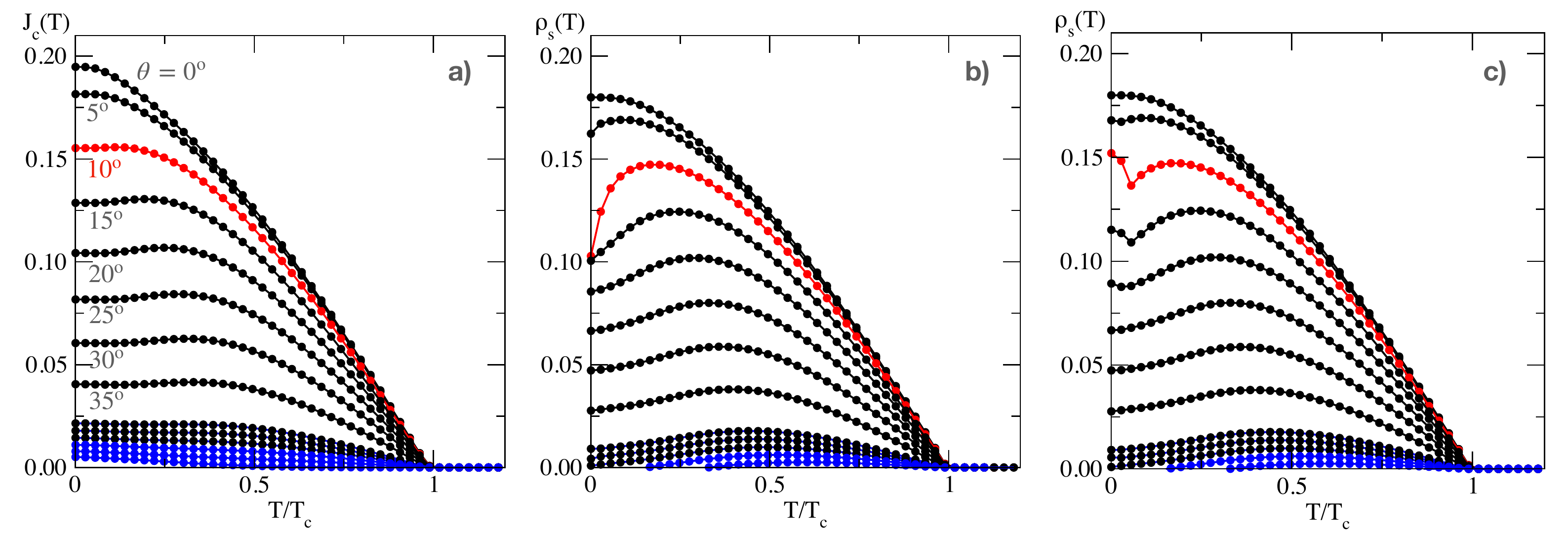}
  \caption{a) Critical current $J_c$ and b) superfluid stiffness $\rho_s$ as a function of temperature for various twist angles. The traces for twist angles above $40^{\rm o}$ are shown in $1^{\rm o}$ increment. Red lines at $\theta=10^{\rm o}$ correspond to the magic angle $\theta_M$. Blue lines show results for $\theta=42-45^{\rm o}$ where time reversal is spontaneously broken at low $T$. Panel c) shows the results for the superfluid stiffness calculated in the presence of the secondary $s$-wave order parameter that arises at low $T$ close to $\theta_M$ as indicated in the phase diagram Fig.\ \ref{fig1}. In these plots the same model parameters are used as in Fig.~\ref{fig3}.}
  \label{fig4}
\end{figure*}

In the following we will be interested in interlayer Josephson current
Eq.\ \eqref{e5} driven by externally imposed phase bias. The
order parameter amplitude $\Delta$ will be determined
self-consistently through the minimization of $\cF_{\rm BdG}$ for a given
phase bias. To this end it is convenient to perform a global gauge
rotation $(c_{\bk 1}, c_{\bk 2})\to (e^{i\varphi/4} c_{\bk 1},
e^{-i\varphi/4} c_{\bk 2})$ which moves the phase factor from the order
parameter terms in $H_\bk$ to the interlayer coupling and 
defines a transformed BdG Hamiltonian 
\begin{equation}\label{h44}
  \tilde{H}_\bk=
 \begin{pmatrix}
   \xi_\bk & \tilde\Delta_{\bk 1} & ge^{-i\varphi/2} & 0 \\
   \tilde\Delta_{\bk 1} & -\xi_\bk & 0 & -ge^{i\varphi/2} \\
   ge^{i\varphi/2}& 0 &   \xi_\bk & \tilde\Delta_{\bk 2} \\
   0 & -ge^{-i\varphi/2} &    \tilde\Delta_{\bk 2} & -\xi_\bk
  \end{pmatrix}.
\end{equation}
Here $\tilde\Delta_{\bk a}$ are defined as in Eqs.\ (\ref{h5},\ref{h6}) but with the phase factors $e^{\pm i\varphi/2}$ omitted. Unless otherwise noted we shall use this representation of the BdG Hamiltonian henceforth, and, for simplicity, we will drop the tilde sign. 

The gap equation follows
from $\partial \cF_{\rm BdG}/\partial\Delta=0$ and reads
\begin{equation}\label{h11}
  \Delta=2\cV\sum_{\bk\alpha}{\partial E_{\bk\alpha}\over \partial
    \Delta}\tanh{{1\over 2}\beta E_{\bk\alpha}},
\end{equation}
where $\beta=1/k_BT$ is the inverse temperature. By noting that
$E_{\bk\alpha}=\bra{\bk\alpha}H_\bk\ket{\bk\alpha}$, where 
$\ket{\bk\alpha}$ is an eigenstate of $H_\bk$, Eq.\ \eqref{h11}
can be rewritten in a form that is more suitable for numerical evaluation, 
\begin{equation}\label{h12}
  \Delta=2\cV\sum_{\bk\alpha}\bra{\bk\alpha}{\partial H_{\bk}\over \partial
    \Delta}\ket{\bk\alpha}\tanh{{1\over 2}\beta E_{\bk\alpha}}.
\end{equation}
Here $\partial H_{\bk}/\partial\Delta$ is a fixed $4\times 4$ matrix that follows from Eq.\ \eqref{h44} and the required matrix element is easily evaluated from the knowledge of the eigenstates. For any chosen phase $\varphi$ and temperature $T$ the gap equation is then solved by iteration starting from a suitable guess for $\Delta$.  

Similarly, by differentiating with respect to $\varphi$ as in Eq.\ \eqref{e5}, it is possible to derive a convenient expression for the interlayer supercurrent
\begin{equation}\label{h13}
  J(\varphi)=-{2e\over \hbar}\sum_{\bk\alpha}\bra{\bk\alpha}{\partial H_{\bk}\over \partial
    \varphi}\ket{\bk\alpha}\tanh{{1\over 2}\beta E_{\bk\alpha}}.
\end{equation}
It is worth noting that this expression is non-perturbative and valid to all orders in the interlayer coupling $g$. This is in contrast to many previous works on twisted junctions \cite{Klemm1998,Klemm2000,Klemm2001} which typically include only the leading $\sim g^2$ term. As we will discuss in greater detail below retaining higher order contributions is crucial for understanding the physics near $\theta=45^\circ$; for example, the double-pair tunneling process that gives the leading non-vanishing contribution in this regime comes in at order $g^4$. 

In the following, we will denote all quantities in units where $2e/\hbar = 1$. For purposes of numerical evaluation the momentum sums in Eqs.\ (\ref{h11}-\ref{h13}) are converted into integrals using the standard procedure summarized in Appendix \ref{app2}, assuming constant density of states $N_F$ and energy cutoff $\epsilon_c$.  


\section{Temperature dependence of the critical current}

\subsection{Current-phase relation} \label{sec:CPR}

In the cuprates individual monolayers are very weakly coupled, implying that $g\ll \mu$ in our model. Correspondingly, the interlayer coupling is a weak perturbation on the SC order parameter amplitude and indeed we find that self-consistently determined $\Delta$ is essentially independent of the phase difference $\varphi$. In the results reported below we use the full phase and temperature dependent $\Delta$ but the same results are obtained if the $\varphi$ dependence were ignored. The temperature dependence, however, cannot be ignored since $\Delta\to 0$ as $T$ approaches $T_c$. 

Typical current-phase relations obtained for parameters relevant to Bi2212 are displayed in Fig.\ \ref{fig3}. At zero twist angle, not surprisingly, we obtain what looks like a conventional sinusoidal $J(\varphi)$ at all temperatures. A fit to Eq.\ \eqref{e6} reveals that a small $J_{c2}$ term is required to capture the data at low $T$. In order to achieve a close fit we add a third harmonic, $-J_{c3}\sin(3\varphi)$. The required $J_{c3}$ is generally two orders of magnitude smaller than $J_{c1,2}$ and has no appreciable effect on the physics. At twist angles close to $45^{\rm o}$, see Fig.\ \ref{fig3}b, a significant second harmonic term is required to capture the form of $J(\varphi)$ obtained from the microscopic model. The negative slope of $J(\varphi)$ at the origin observed for the low $T$ curves signals the spontaneously $\cT$-broken phase: $\varphi=0$ corresponds to a local maximum of $\cF(\varphi)$ while two minima occur at a non-zero phase $\pm\varphi_{\rm min}$. 

Panel (c) shows an example of the dependence of parameters $J_{c\alpha}$ with $\alpha=1,2,3$ on temperature, typical for twist angles close to $45^{\rm o}$. We observe that at low $T$ the second harmonic $J_{c2}$ dominates over the regular Josephson tunneling $J_{c1}$. 
By differentiating Eq.\ \eqref{e6} it is easy to see that the condition $\rho_s=0$ is attained when $J_{c1}=2 J_{c2}$, which implies that the
system is in the $\cT$-broken phase when $J_{c1}<2 J_{c2}$. This is the criterion we used to establish the phase diagram shown in Fig.\ \ref{fig1}.


\subsection{Critical current and superfluid density}

Fig.\ \ref{fig4} summarizes our results for the $T$-dependent interlayer critical current $J_c(T)$ and superfluid stiffness $\rho_s(T)$ defined in Eqs.~(\ref{e7},\ref{e8}). Several interesting features can be seen. At intermediate twist angles $J_c$ exhibits an {\em increase} with increasing temperature (see panel a). Such behavior is anomalous -- normally thermal excitations cause Cooper pair breaking which depletes the SC condensate and results in a monotonic decrease of $J_c$ as a function of temperature -- and it reflects the sign-changing nature of the $d$-wave order parameter and provides an experimentally accessible signature of unconventional superconductivity in twisted cuprate bilayers. We note that $\rho_s$ shows similar behavior which is even more pronounced, especially near the magic angle $\theta_M\simeq 15^{\rm o}$.  

The anomalous increase in $\rho_s(T)$ and $J_c(T)$ can be understood as follows. We focus on $\rho_s(T)$ but similar arguments apply to $J_c(T)$. As shown in the Appendix \ref{app1} the superfluid stiffness can be expressed as 
\begin{equation}\label{s1}
\rho_s(T)=\sum_\bk \Delta_{\bk 1}\Delta_{\bk 2}\Omega(\xi_\bk,T),
\end{equation}
where $\Omega(\xi_\bk,T)\geq 0$. The key observation is that the product of the two gap functions $\Delta_{\bk 1}\Delta_{\bk 2} = \Delta^2 \cos(2\alpha_\bk - \theta) \cos(2\alpha_\bk + \theta)$ becomes negative for angles $\alpha_\bk$ close to $\pm \pi/4$ when the twist angle is non-zero. At $T=0$ these nodal regions of $k$-space therefore make a {\em negative} contribution to $\rho_s$. This negative contribution grows with increasing twist which accounts for decreasing $\rho_s(0)$ observed in Fig.\ \ref{fig4}b. At nonzero temperature thermal excitations cause Cooper pair breaking and at low $T$ this happens predominantly in the nodal regions of $k$-space where the low-lying excitations reside. Since Cooper pairs in the nodal regions make negative contribution to $\rho_s$, removing these leads to an anomalous {\em increase} of $\rho_s(T)$ with increasing temperature apparent in Fig.\ \ref{fig4}. 

The $\cT$-broken phase is signalled by $\rho_s$ turning negative. For our chosen parameters this occurs when $\theta>42^{\rm o}$ and corresponds to blue curves in Fig.\ \ref{fig4}. Unfortunately, while straightforward to calculate theoretically, $\rho_s$ is not easily measurable. $J_c$ on the other hand, which is experimentally accessible, does not show any clear signatures of the spontaneously $\cT$-broken phase. Physically, $J_c$ represents the largest current that the system can sustain before going normal. At such a large externally imposed current the time-reversal is explicitly and strongly broken, so $J_c$ will not be sensitive to a small amount of $\cT$-breaking already present in the unbiased system. On the other hand, $\rho_s$ is defined at $\varphi=0$ which corresponds to a vanishing supercurrent and there is no explicit $\cT$-breaking. Therefore, it is natural to expect that temperature dependence of $\rho_s$ can serve as a sensitive probe of spontaneous $\cT$ breaking.  


\subsection{Estimate of interlayer coupling $g$}

The strength of the interlayer coupling $g$ is a key parameter that, in a twisted bilayer, determines the position of the magic angle $\theta_M$ as well as the width of the $\cT$-broken topological phase near the $45^{\rm o}$ twist. In the literature on high-$T_c$ cuprates one can find many estimates of $g$ but the values vary widely for different materials. For Bi2212 one finds values, obtained from both experimental fits and theoretical modelling, ranging between $5-120$ meV depending on the technique used (see e.g.\ Ref.\ \cite{Can2021} for a recent summary). This wide range can be attributed to the complicated crystal and electronic structure of Bi2212, as well as the fact that $g$ is likely momentum-dependent and different probes are sensitive to different parts of the Brillouin zone. 

To be able to provide quantitative predictions for transport in twisted Bi2212 bilayers, we now attempt to extract the likely value of $g$ relevant to the Josephson phenomena discussed in this paper by comparing our theoretical results to the recent critical current measurements in twisted Bi2212 flakes \cite{Zhao2021}. Given various uncertainties we give two separate estimates; one based on the GL theory valid at high temperatures and one based on microscopic modeling relevant in the $T\to 0$ limit.

The experimental measurement (Fig.\ 2D in Ref.\ \cite{Zhao2021}) gives temperature dependence of $J_c R_N$ for various twist angles. Here $R_N$ is the normal-state junction resistance which serves to normalize $J_c$ for different junctions, effectively providing a measure of the critical current density that is independent of the junction area and geometry. Our GL estimate relies on extracting the critical angle $\theta_c$ from Eq.\ \eqref{e88} based on the data and then backing out the value of $g$ by matching to the microscopic phase diagram shown in Fig.\ \ref{fig1}. We chose $T=50$ K as our reference temperature and read off $J_cR_N=(10.4,8.2,3.5,0.2)$ mV at $\theta=(0^{\rm o},29^{\rm o},39^{\rm o},44.9^{\rm o})$. We assume that for twist angles $\theta< 40^{\rm o}$ the current-phase relation \eqref{e9} is dominated by the first term and we can approximate $J_c(\theta)\simeq J_{c1}(\theta)=J_{c1}(0)\cos(2\theta)$. Close to $45^{\rm o}$ the second term dominates and we take $J_{c2}\simeq J_c(44.9^{\rm o})$. Eq.\ \eqref{e88} then gives $\theta_c=(44.1\pm0.2)^{\rm o}$. Finally, comparing to the microscopic phase diagram in Fig.\ \ref{fig1} we see that this range of critical angles at $T=50$ K is well captured by taking $g\simeq 10.5$ meV. This corresponds to $\theta_c\simeq 39^\circ$ in the $T\to 0$ limit and $\theta_M\simeq 10^\circ$.

A more direct but technically somewhat more challenging estimate of $g$ can be given starting from the microscopic expression for $J(\varphi)$ given in Eq.\ \eqref{a2} of Appendix \ref{app1}. Because at $\theta=0$ the current is dominated by the ordinary Josephson tunneling while near $45^{\rm o}$ double Cooper pair tunneling dominates, one expects the leading term in the $T\to 0$ limit to behave in the two cases as $\sim g^2$ and $\sim g^4$, respectively. Indeed as shown in Appendix \ref{app3} one can estimate, at $T=0$, 
\begin{equation}\label{s2}
J_c(\theta=0^{\rm o})\simeq 2\pi N_F{eg^2\over \hbar}C_2,
\end{equation}
while 
\begin{equation}\label{s3}
J_c(\theta=45^{\rm o})\simeq 2\pi N_F{eg^4\over \hbar\Delta^2}C_4.
\end{equation}
Here $C_2$ and $C_4$ are dimensionless constants of order one which depend on model details such as the Fermi surface shape and the pairing interaction cutoff scale $\epsilon_c$. For simplicity and concreteness we assume a circular Fermi surface and work in the limit $\Delta\ll\epsilon_c$. Under these conditions $C_2=1$ and $C_4\simeq 0.55$. Taking a ratio one can determine $g$ as
\begin{equation}\label{s4}
g\simeq \Delta\sqrt{{C_2\over C_4}{J_c(\theta=45^{\rm o})\over J_c(\theta=0^{\rm o})}}.
\end{equation}
From the experimental data (Fig.\ 2D in Ref.\ \cite{Zhao2021}) we estimate $J_cR_N(\theta=45^{\rm o})/ J_cR_N(\theta=0^{\rm o})\simeq 0.010$, which together with $\Delta=45$ meV gives $g\simeq 6.2$ meV.

Given various simplifying assumptions, the two estimates agree reasonably well. In view of the values found in the literature (typically 10s of meV), we conclude that $g$ around 10 meV is likely the most appropriate value to use when modelling Josephson effects in twisted Bi2212.


\subsection{Lattice model and dependence of critical current on interlayer coupling}

\begin{figure}
    \centering
    \includegraphics[width=8cm]{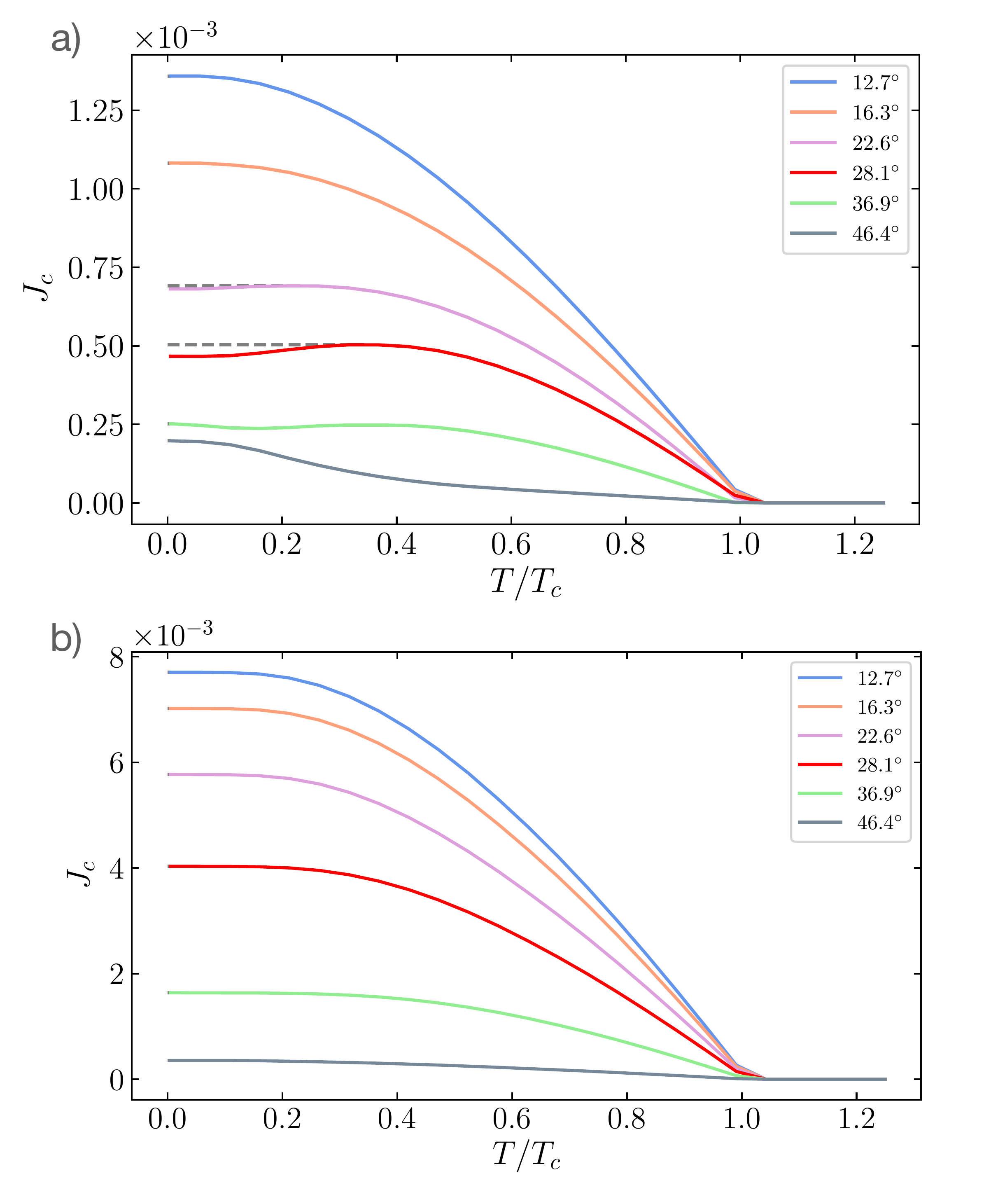}
    \caption{Behavior of the critical current for various accessible commensurate twist angles in the lattice model. For a detailed discussion of the model, see Ref.\ \cite{Can2021}. The parameters used are $\mu = -1.3t$, $g_0=10$meV in a) and $\mu = -1.4t$, $g_0=30$meV in b).}
    \label{fig5}
\end{figure}

The results reported in Fig.\ \ref{fig4} were obtained using a simple continuum model for a $d$-wave SC. Because the interesting low-$T$ behavior arises from the physics of the nodal excitation which are well described by the continuum model, we expect this model to provide an accurate description of twisted bilayer cuprates at low $T$. To ascertain the robustness of these results, we have additionally computed the critical current from a more realistic lattice model defined in Ref.\ \cite{Can2021}, which accurately captures the hole-like Fermi surface of Bi2212 near optimal doping. Fig.\ \ref{fig5} shows the temperature dependence of the critical current calculated for a range of commensurate angles $\theta_{m,n}=2 \arctan(m/n)$ with integers $m,n$. For these angles the twisted bilayer forms a Moire pattern with $2(m^2+n^2)$ sites per unit cell.

As in the continuum model, an anomalous increase in $J_c$ at low temperatures for intermediate twists is observed in Fig.\ \ref{fig5}a. Such behavior is observed only if the interlayer tunneling is small. Specifically, as shown in Fig.\ \ref{fig5}b, when $g$ is increased the curves resemble the familiar Ambegaokar-Baratoff form \cite{Ambegaokar_1963} where the current monotonically decreases with $T$. Note that this dependence on tunneling strength is a feature also of the continuum model discussed in the previous sections. We attribute this change in behavior to non-perturbative effects: when $g$ becomes comparable to $\Delta$ simple perturbative arguments underlying our reasoning in Sec.\ III.B no longer apply. Nevertheless the observed agreement in the characteristic behavior of the two models lends support to our predictions for $J_c(T)$ in twisted cuprates.


\section{Secondary order parameters nucleated near the magic angle} 

An interesting situation occurs at small twist angles, when, as a function of increasing $\theta$, two Dirac points originating from the two layers collide and form a quadratic band crossing (QBC). As discussed in Ref.\ \cite{volkov2020magic}, such a collision is unavoidable as long as $\cT$ and two-fold rotation symmetry along the in-plane $y$ axis are respected. The QBC then occurs at a magic angle $\theta_M$ and is analogous to the flat-band formation in twisted bilayer graphene. 

Similar to bilayer graphene, the system becomes susceptible to interactions for twist angles close to $\theta_M$. For two separate Dirac points the density of states (DOS) tends to zero at zero energy and interactions are perturbatively irrelevant. At a QBC, however, the DOS becomes constant at low energies and the system behaves, essentially, like a metal of Bogoliubov quasiparticles. Such a metal will be unstable with respect to residual interactions that have not been included in the original mean-field treatment of $d$-wave superconductivity. For instance, as discussed in Ref.\ \cite{volkov2020magic}, if there exist attractive interactions in the $s$ or $d_{xy}$ channel, the system will develop a secondary SC order parameter with that symmetry for twist angles close to $\theta_M$. Furthermore, because the spectrum is gapless, this will happen for arbitrarily weak attractive potential.

To illustrate this behavior we focus on the secondary instability in the $s$-wave channel and investigate its effect on the behavior of $J_c(T)$ and $\rho_s(T)$. Working in the gauge described below Eq.\ \eqref{h8} the pair amplitudes can be expressed as
\begin{eqnarray}
\Delta_{\bk 1}&=& \Delta \cos(2\alpha_\bk-\theta)+i\Delta_s, \\
\Delta_{\bk 2}&=& \Delta \cos(2\alpha_\bk+\theta)+i\Delta_s,        
\end{eqnarray}
where $\Delta$ and $\Delta_s$ are assumed real. The imaginary unit in front of $\Delta_s$ is required to break $\cT$; the QBC is protected by $\cT$ and a gap can only open when time reversal is broken. The gap equation for the secondary order parameter follows from  minimizing the free energy Eq.\ \eqref{h8} with respect to $\Delta_s$ and reads
\begin{equation}\label{g2}
  \Delta_s=2\cV_s\sum_{\bk\alpha}\bra{\bk\alpha}{\partial H_{\bk}\over \partial
    \Delta_s}\ket{\bk\alpha}\tanh{{1\over 2}\beta E_{\bk\alpha}},
\end{equation}
where $\cV_s$ denotes interaction strength in the $s$ channel.

Fig.\ \ref{fig1} shows the typical phase diagram obtained by numerically solving the coupled gap equations \eqref{h12} and \eqref{g2}. We observe a broad dome of $s$ order parameter peaked at $\theta=\theta_M$ which occurs in addition to the previously discussed $d+id'$ phase. It is to be noted that parameters in Fig.\ \ref{fig1} are chosen so as to maximize the secondary instability; if $\cV_s$ were chosen any larger we would find nonzero $\Delta_s$ even at zero twist angle, contrary to experimental observations which indicate pure $d_{x^2-y^2}$ order parameter in untwisted crystals and films. These considerations suggest that generically secondary order parameter physics will be only visible at low temperatures compared to the native $T_c$ of the cuprates. We also note that similar results are obtained for a secondary instability in the $d_{xy}$ channel. 

$J_c(T)$ and $\rho_s(T)$ can be calculated in the presence of the secondary order parameter as before. We find that $\rho_s(T)$ shows a clear signature of the secondary order illustrated in Fig.\ \ref{fig4}c. The enhancement at low $T$ can be attributed to the extra contribution of the $s$-wave Cooper pairs to the tunneling between the layers. $J_c(T)$, on the other hand, is unaffected by the secondary order parameter. This is because critical current is achieved when the phase difference $\varphi$ is close to $\pi/2$; we find that in this regime Eqs.\ \eqref{h12} and \eqref{g2} produce self-consistent solution with $\Delta_s$ strongly suppressed compared to its value at zero phase. Thus, one more time we find that the critical current alone is not a suitable probe for establishing the presence of the secondary order parameters.


\section{Fraunhofer pattern} \label{sec:fraunhofer}

Fraunhofer interference occurs when a Josephson junction is subjected to an external magnetic field parallel to the junction plane \cite{tinkham2004introduction,Barone_1982}. 
To describe the effect of such magnetic field on the inter-plane Josephson phase and current distribution we must allow for spatial variation of the order parameters $\psi_a$ in the plane. The starting point of the calculation is as in Eq.~\eqref{e1}, except that the interlayer Cooper pair hopping terms now contain magnetic phases picked up when tunneling between the layers. We thus consider a free energy {\em density}
\begin{eqnarray}\label{f1}
 f[\psi_1,\psi_2]&=&f_0[\psi_1]+f_0[\psi_2]+A |\psi_1|^2
                      |\psi_2|^2\\
  &+&B(\psi_1\psi_2^\ast e^{-iqx} +{\rm c.c.})+C(\psi_1^2\psi_2^{\ast 2} e^{-2iqx} +{\rm c.c.}).\nonumber
\end{eqnarray}
where $q=Bd/\Phi_0$ encodes the magnetic flux per unit length that is threaded between the layers separated by distance $d$, and 
\begin{equation}\label{f2}
f_0[\psi]=\alpha|\psi|^2+{\frac{1}{2}}\beta|\psi|^4+\gamma|\nabla\psi|^2
\end{equation}
describes each layer. We choose a gauge $\psi_2 = \psi e^{i(\varphi_x -qx)}$ in Eq.~\eqref{f1}, where $\varphi_x$ is the spatially dependent Josephson phase difference and $\psi$ is assumed constant and real. This leads to a free energy density
\begin{eqnarray}\label{eq:Fdensity}
f(\varphi_x) = f_0 &+& \tilde{\gamma} (\partial_x \varphi_x - q)^2 \\ \notag
&-& \frac{\hbar}{2eV}\left[
J_{c1}\cos\varphi_x - \frac{J_{c2}}{2}\cos(2\varphi_x)\right],
\end{eqnarray}
where $V$ denotes the 2D volume of the system, $\tilde\gamma=\gamma\psi^2$, and we omitted higher harmonic contributions ($J_{c3}$) that are found to be very small, see Fig.~\ref{fig3}.

We observe that in addition to the parameters $J_{c\alpha}$, one also requires the in-plane superfluid stiffness, related to the parameter $\gamma$ in the free energy Eq.~\eqref{f2}. The calculation of this quantity for a single-layer $d$SC is standard in the high-$T_c$ literature \cite{Scalapino1993,Hirschfeld1993,Wang2001,Sheehy2004}, but for the sake of completeness we include it for our specific model in Appendix \ref{app2}. Using this input, one can obtain results for the phase evolution and winding of the SC phase difference $\varphi(x) = \varphi_x$ as a function of both the applied in-plane magnetic flux and temperature.

\begin{figure}
    \centering
    \includegraphics[width=\columnwidth]{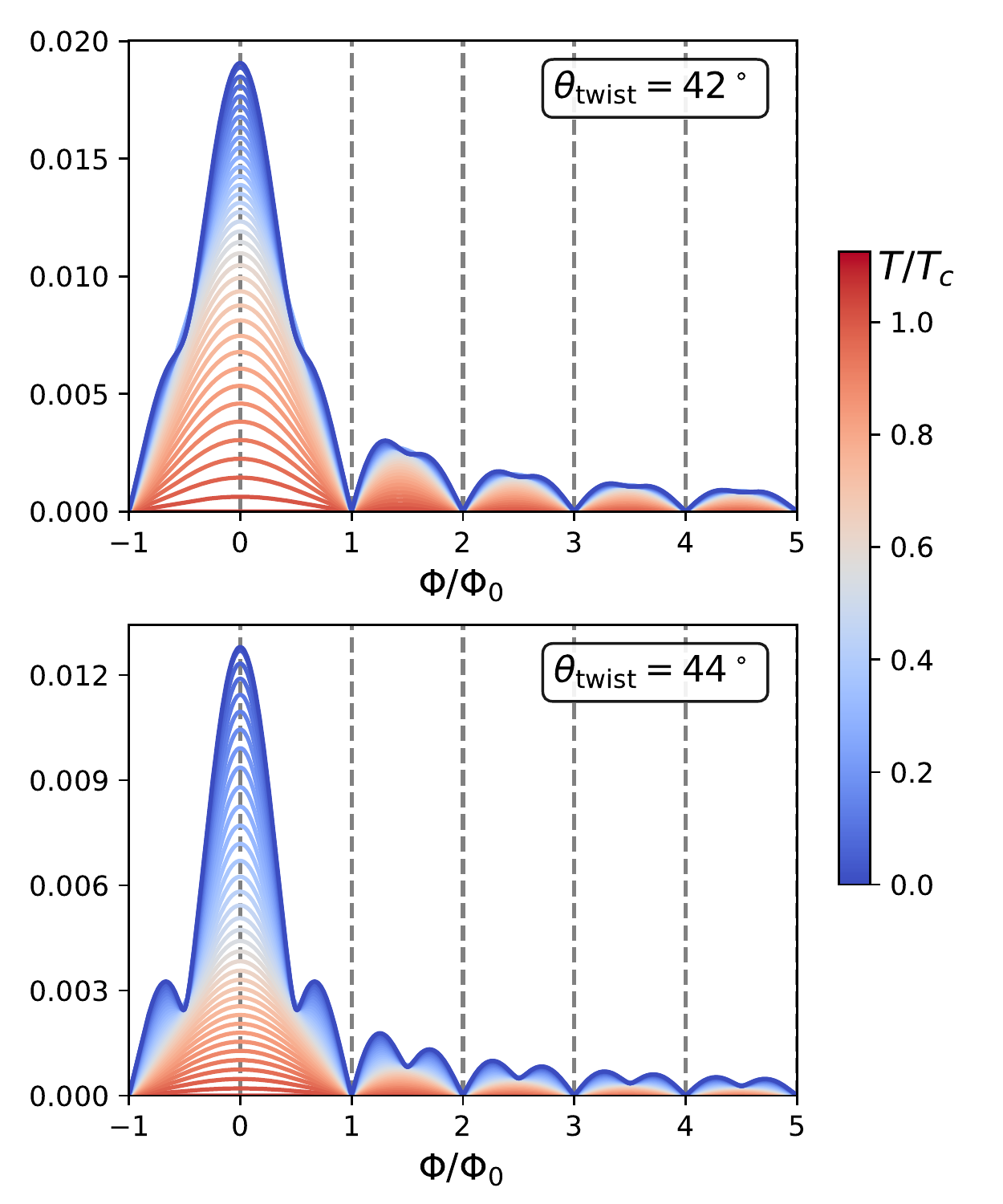}
    \caption{
    Critical current $I_c(\Phi,T)$ vs applied magnetic flux $\Phi$ in the twist junction for various temperatures $T$, with twist angle $\theta = 42^\circ$ (top) and $\theta = 44^\circ$ (bottom). The curves are computed from a Lawrence-Doniach type model of the bilayer junction, taking into account the first and second harmonic terms of the current in Eqs.~\eqref{e6},~\eqref{eq:jx} with coefficients $J_{c1,c2}$ fitted from Fig.~\ref{fig3}, closely following the discussion of Ref.~\cite{Goldobin_2007}.}
    \label{fig:fraunhofer}
\end{figure} 
For a set of zero-field model parameters $\gamma$ and $J_{c1,2}$, it is easy to discretize and numerically determine solutions $\varphi_x$ that minimize the above free energy for a given total flux $\Phi = qW$ in a twist junction of width $W$. For small to moderate flux densities (magnetic fields), we find that the local Josephson phase increases approximately linear, $\varphi_x \approx \varphi_0 + qx$. This is expected in short Josephson junctions~\cite{Bulaevskii_1992,Goldobin_2007}, trivially minimizes the kinetic term, and leaves one with the (local) Josephson current-phase relation,
\begin{equation}\label{eq:jx}
j_x(\varphi_x) = j_{c1}\sin(\varphi_0+qx) - j_{c2}\sin(2\varphi_0+2qx)~,
\end{equation}
where $j_{ca}=J_{ca}/V$ denote the respective densities. 
We then calculate the critical current by integrating the above Josephson current density and maximizing with respect to the phase offset $\varphi_0$, as described by Goldobin et al.~\cite{Goldobin_2007}. The final expression for the total critical current (normalized by $V$) then takes the form
\begin{equation}\label{eq:fraunhofer}
I_c(\theta,\Phi,T) = 
I_{c1}\left|{\sin(\pi\Phi/\Phi_0)\over\pi\Phi/\Phi_0}\right| +
I_{c2}\left|{\sin(2\pi\Phi/\Phi_0)\over 2\pi\Phi/\Phi_0}\right|,
\end{equation}
where $I_{c\alpha} = I_{c\alpha}(\theta, T) \sim J_{c\alpha}(\theta, T)$.
As already alluded to in the Introduction Eq.~\eqref{e9}, the Fraunhofer pattern thus interpolates between regular and period-halved cases, as a function of both twist angle and temperature. 

We show data for twist angles $\theta = 42^\circ$ and $44^\circ$ and with varying temperature $T$ in Fig.~\ref{fig:fraunhofer}. Since the $d+id'$ phase is prominent only for a range of twist angles around $\theta = 45^\circ$, the period-halved contribution in the Fraunhofer pattern is rapidly lost when moving to $\theta \lesssim 40^\circ$. In this case the pattern comprises slightly deformed regular periodic lobes which might be difficult to identify experimentally. However as $\theta \to 45^\circ$ the period-halved contribution becomes dominant, see Fig.~\ref{fig:fraunhofer} bottom panel. With the presently accessible twist angle control, and assuming that samples have a fairly uniform twist angle distribution, it should thus be feasible to observe period-halved Fraunhofer patterns in experiment.
Finally, we note that while the precise magnitudes and dependencies of parameters $I_{c1,2}(\theta,T)$ in Eq.~\eqref{eq:fraunhofer} depend on the assumptions made in this section (e.g.\ linear growth of $\varphi_x$), the functional form with first and second harmonics in the Fraunhofer pattern is generic.



\section{Shapiro steps} \label{sec:shapiro}

\begin{figure*}[t]
  \includegraphics[width=18cm]{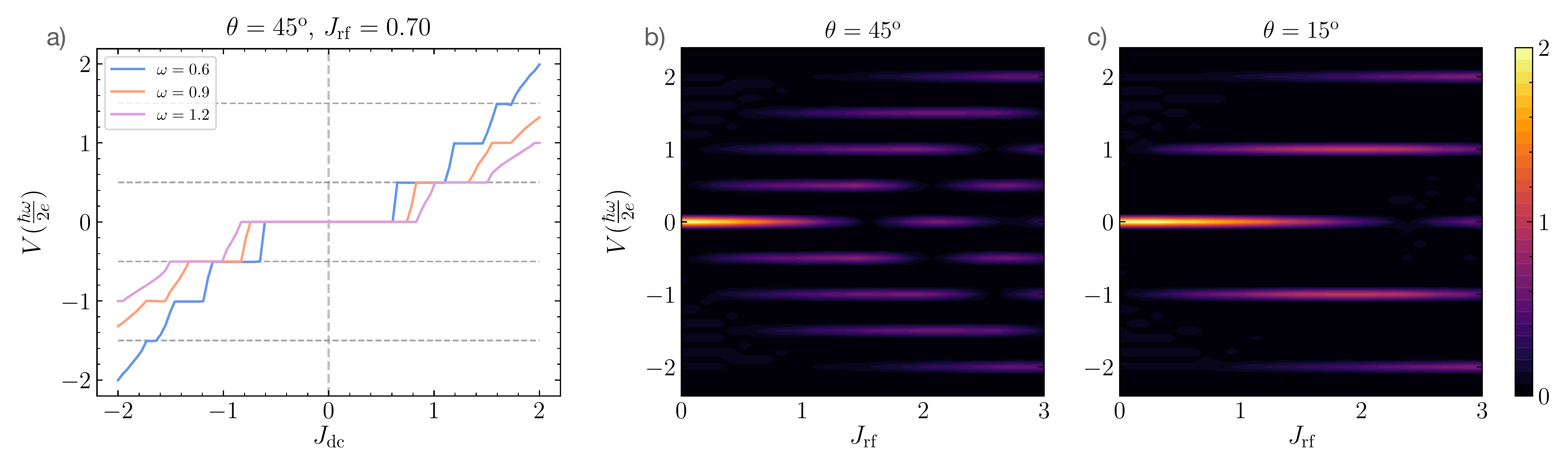}
  \caption{a) Current-Voltage curves in the $45^{\rm o}$ twisted configuration in the presence of an external rf drive with amplitude $J_{\rm{rf}}=0.7$ at three different frequencies. The voltage is scaled in units of $\hbar \omega / 2e$ to highlight the Shapiro steps and the lowest four fractional steps are indicated with the horizontal dashed lines. b-c) Voltage as a function of drive current for two twists that are in the topological and trivial phases respectively, with $\omega=0.6$. The color bar indicates step amplitudes. Since the numerical values of $J_{c1,2}$ at different twists vary widely, to aid comparison, $J(\varphi)$ is normalized such that the critical current in the absence of rf drive ($J_{\rm{rf}}=0$) is unity (or equivalently the zeroth order step has width 2). Clearly, fractional steps appear only for twist angles close to $45^\circ$, as in panel b). At high rf currents, one enters a regime where the step amplitudes oscillate. The temperature is fixed at $T = 0.5T_c$ and $R=0.7$ in all simulations.}
  \label{fig:shapiro}
\end{figure*} 
Topological superconductivity with a dominant second harmonic in the current-phase relation can also be made manifest by subjecting the system to an external drive. If a Josephson junction is irradiated with a radio-frequency source, an AC voltage is induced that frequency modulates the AC Josephson current. When the dynamics of the junction is phase-locked to the rf drive, the supercurrent shows constant voltage Shapiro steps at voltages $V_n = n\hbar \omega / 2e$, where $\omega$ is frequency of the radiation and $n \in \mathbb{Z}$ is the step index \cite{Shapiro_1963}. In a $\pi$ periodic junction, additional steps appear at fractional values $V_{n/2}$, which are a direct manifestation of a dominant second harmonic in the current-phase relation.

In a usual experiment, the time-averaged voltage is measured in response to a current bias. In order to study the dynamics of such a current driven system, it is convenient to work within a semiclassical framework where current is carried in three parallel channels: supercurrent carried by the Cooper pairs, a resistive path for dissipative current and a capacitive channel that accounts for charge build up on the superconducting leads. If the junction has a negligible geometric capacitance, one is in an over-damped regime where the resistive shunt provides the only impedance. In this so-called resistively shunted junction (RSJ) model \cite{Barone_1982}, the total current through the system can be written as $J = J_{R} + J(\varphi)$, where $J_R = V/R$ is the current through the resistor $R$ and $J(\varphi)$ is the current-phase relation \eqref{e6}. Finally, utilizing the universally valid superconducting phase evolution relation $d\varphi / dt = 2eV/\hbar$ and imposing the current drive $J = J_{\rm{dc}} + J_{\rm{rf}} \sin(\omega t)$, one obtains the first order differential equation 

\begin{equation}
    \frac{\hbar}{2 e R} \frac{\partial \varphi}{\partial t} + J(\varphi) = J_{\rm{dc}} + J_{\rm{rf}} \sin(\omega t).
    \label{eq:rsj}
\end{equation}

For a given twist, temperature, rf-drive parameters, $R$ and $J_{\rm{dc}}$, the time evolution of the phase and, via a time derivative, the voltage can be obtained by solving the above equation numerically through a routine Runge-Kutta algorithm. The representative time-averaged voltage behavior as a function of the direct current is depicted in Fig.\,\ref{fig:shapiro}a, where fractional steps corresponding to the halved period in the $\cT$-broken phase are seen. Another way to visualize the Shapiro physics is to study the dependence of the step widths as a function of rf current, where the steps are revealed as maxima at quantized voltages, see Fig.\,\ref{fig:shapiro}b-c. Since $P_{\rm{rf}}=J_{\rm{rf}}^2 R_{\rm{rf}}$, $J_{\rm{rf}}$ also serves as a proxy for rf power. Steps appear progressively, starting with low values of $n$, as the drive amplitude is increased  and at higher powers one observes an oscillatory pattern \cite{Russer_1972, tinkham2004introduction}.

The constant voltage steps may be intuitively understood as the virtual tunneling of Cooper pairs across the barrier that is accompanied by an exchange of photons with the radiation. Whenever the potential energy across the junction is equal to the photon energy $\hbar \omega$, or a multiple thereof, a Cooper pair can absorb (emit) photons from (to) the radiation field. The $n^{\rm th}$ Shapiro step corresponds to $n$ photons being exchanged. The fractional steps, on the other hand, coincide with two Cooper pairs tunnelling across the junction.

With all else fixed, step amplitudes are proportional to the critical currents $J_{c1, 2}$ that are informed by the microscopic model. To observe all steps and avoid interference, one has to ensure that they are sufficiently apart on the $J_{\rm{dc}}$ axis; the separation is controlled by the amplitude and frequency of the drive. While $J_{c2}$ is non-zero at any point in the phase diagram, we find empirically that having $J_{c2}/J_{c1} > 2$ is required to discern the fractional steps. This is a stronger condition than  $J_{c2}/J_{c1} > 1/2$ which, according to our discussion in Sec. \ref{sec:CPR}, defines the $\cT$-broken phase. Therefore, appearance of fractional Shapiro steps in experimental data can be regarded as a strong sign of the $\cT$-broken phase in the system.
\begin{figure*}[t]
  \includegraphics[width=16cm]{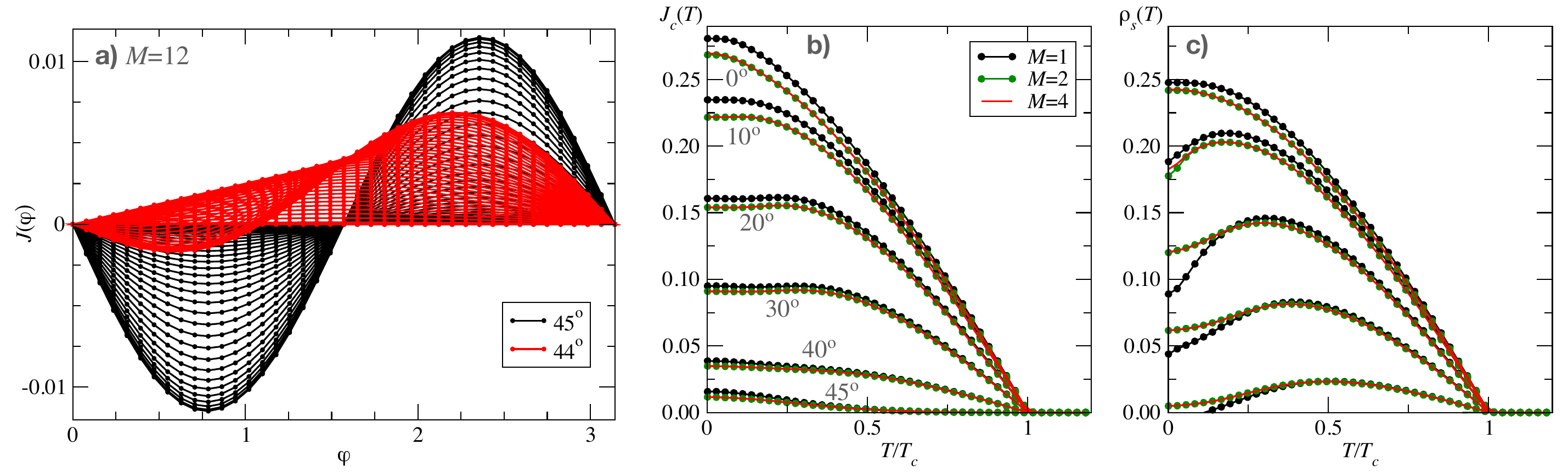}
  \caption{a) Current-phase relation computed for two flakes, each with $M=12$ monolayers, for twist angles near $45^{\rm o}$. Various curves are for temperatures $T$ between zero and $T_c$ in increments $T_c/20$. b) Critical current $J_c(T)$ and c) phase stiffness $\rho_s(T)$ extracted from the calculated $J(\varphi)$. We obtained results for $M$ up to 24 but we find that the curves become essentially independent of $M$ when $M\geq 4$. }
  \label{fig10}
\end{figure*} 

Temperature implicitly enters our model through microscopic parameters $J_{c1, 2}$. To fully account for the thermal effects, one needs to add a noise term to the bias current \cite{Ambegaokar_1969}. The main consequence of such a treatment is the `rounding’ of steps in the current-voltage characteristic: In the presence of fluctuations, the switch from a constant voltage plateau to a dissipative state would no longer be sharp, resulting in smaller step amplitudes. Nevertheless, noise does not alter the location of the steps and the qualitative features discussed above continue to hold.

It is to be noted that fractional Shapiro steps can arise in physical systems that are removed from the context of topological superconductivity. For instance, two-dimensional Josephson junction arrays may show half-integer steps due to a skewed current-phase relation \cite{Benz_1990, Panghotra_2020}. A $\sin(2 \varphi)$ dependence of the current also appears in magnetic Josephson junctions \cite{Sellier_2004}. In the present setting, given our theoretical understanding, topological superconductivity should be the primary candidate for the physics underlying unconventional Shapiro physics.

\section{Thicker flakes}

In this section we briefly consider twisted structures composed of thicker flakes that might be easier to assemble and probe in the lab~\cite{Zhao2021,Xue2021}. Specifically, we study the Josephson current between two flakes each composed of $M$ cuprate monolayers. The microscopic Hamiltonian describing this situation can be constructed as a straightforward extension of Eq.\ \eqref{h4}. For example the  $M=2$ system is represented by an $8\times 8$ matrix BdG Hamiltonian 
\begin{equation}\label{t1}
    H_\bk=\begin{pmatrix}
      h_{\bk 1} & \Gamma & 0 & 0 \\
      \Gamma & h_{\bk 1} & \Gamma & 0 \\
      0 & \Gamma & h_{\bk 2} & \Gamma \\
      0 & 0 & \Gamma & h_{\bk 2} 
    \end{pmatrix},
\end{equation}
where 
\begin{equation}\label{t2}
    h_{\bk a}=\begin{pmatrix}
\xi_\bk & \Delta_{\bk a} \\
\Delta_{\bk a}^\ast & -\xi_\bk   
    \end{pmatrix}, \ \ \ \ \
   \Gamma=\begin{pmatrix} 
   g & 0 \\
   0 & -g
   \end{pmatrix},
\end{equation}
describe the individual monolayers and their coupling, respectively. The twist between the flakes is encoded in the structure of the order parameter given in Eqs.\ (\ref{h5},\ref{h6}). For a generic
$M$ the Hamiltonian becomes a $4M\times 4M$ matrix with a structure that follows as an obvious generalization of Eq.\ \eqref{t1}.

The supercurrent flowing between the two flakes can be calculated from Eq.\ \eqref{h13}. On this basis we find that many of the features observed previously for two coupled monolayers (the $M=1$ case) persist in thicker flakes. Fig.\ \ref{fig10} shows some representative results.  In panel (a) we observe that the characteristic period doubling in $J(\varphi)$ as the twist approaches $45^{\rm o}$ persists for $M>1$. The critical current $J_c$ and phase stiffness $\rho_s$ likewise show similar characteristic temperature dependence for $M>1$ as monolayer devices, except for a slight suppression of the amplitude at low $T$. Interestingly, the largest change occurs between $M=1$ and $M=2$ cases; for $M\geq 2$ the system appears to have reached the bulk limit and we see no perceptible change in $J_c(T)$ and $\rho_s(T)$ as $M$ is increased further.

\begin{figure}[t]
  \includegraphics[width=8.6cm]{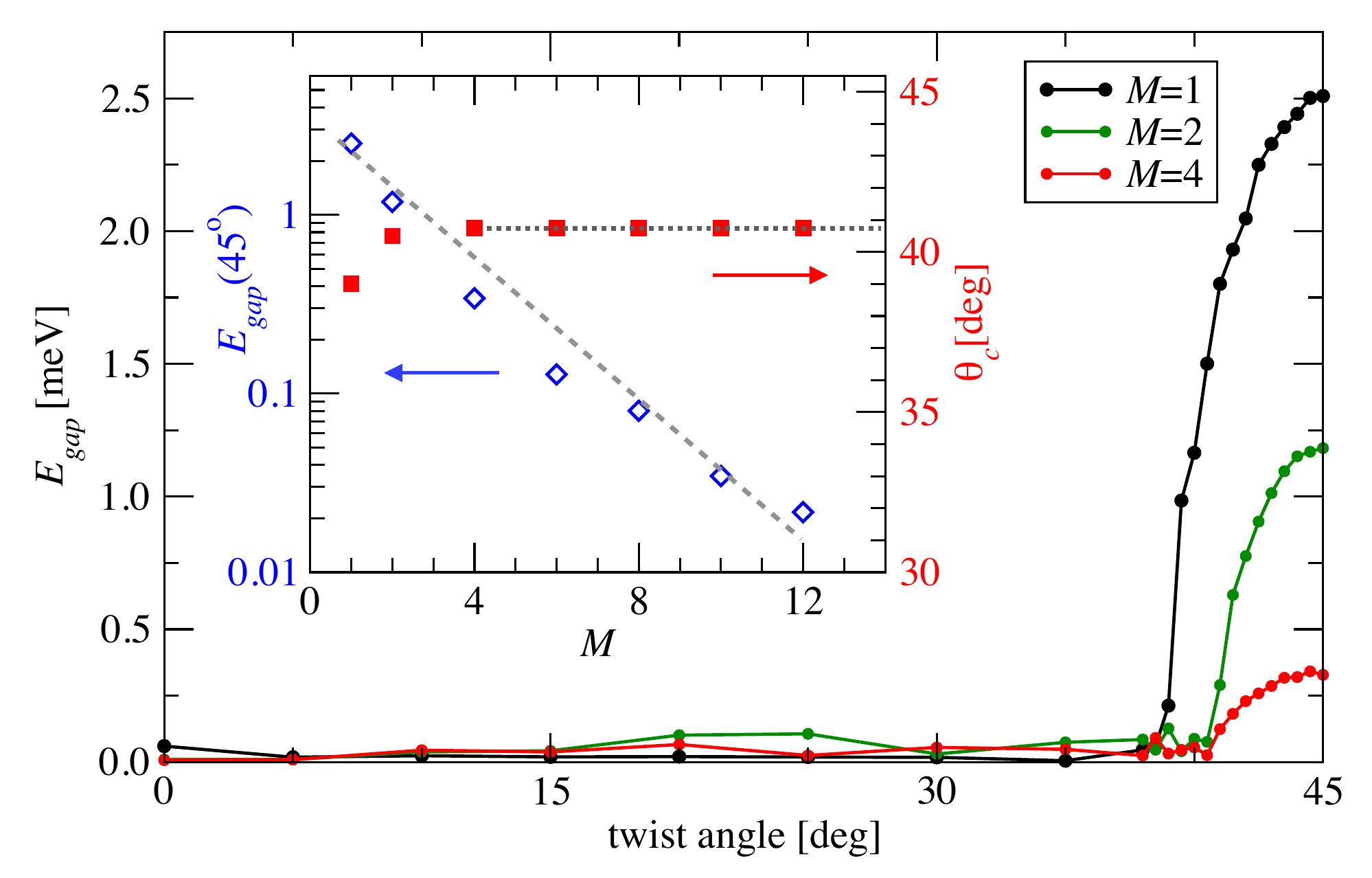}
  \caption{Evolution of the spectral $E_{\rm gap}$ with the flake thickness $M$. Note that $E_{\rm gap}$ values below $\sim 0.1$ meV are comparable to the finite-size gap in our system ($500\times 500$ grid of $\bk$ points) and are therefore consistent with gapless behavior. Inset shows scaling of the maximum gap $E_{\rm gap}(45^{\rm o})$ with $M$, suggesting exponential decay, and the critical twist angle $\theta_c$ beyond which the $\cT$-broken phase sets in.}
  \label{fig11}
\end{figure}

A quantity that strongly depends on thickness $M$ is the spectral gap $E_{\rm gap}$ which we define as the energy of the lowest quasiparticle excitation  above the ground state, obtained as the smallest positive eigenvalue of $H_\bk$. The spectral gap is measurable through various spectroscopic probes, such as electron tunneling or angle-resolved photoemission.  Fig.\ \ref{fig11} shows our results for the spectral gap as a function of the twist angle for several values of $M$. The gap is largest at $\theta=45^{\rm o}$, where the $\cT$-breaking induced by the interlayer tunneling is expected to be maximal. The maximum gap $E_{\rm gap}(45^{\rm o})$ is seen to rapidly decay with increasing $M$; the inset suggests an exponential dependence on $M$. Indeed this is to be expected: in the limit of thick flakes the tunneling becomes a surface perturbation to a 3D system which cannot open a bulk gap. Nevertheless the $\cT$ broken phase persists in this limit as can be seen from the behavior of the critical twist angle $\theta_c$ also shown in the inset. $\theta_c$ is defined as the twist angle at which the free energy $\cF_{\rm BdG}$ first develops minima away from $\varphi=0$, signalling the onset of spontaneous $\cT$ breaking. The inset shows that with the increasing flake thickness $\theta_c$ quickly approaches saturation, leading to a stable range of twist angles $|\theta-45^{\rm o}|<\theta_c$ in which the system ground state exhibits spontaneously broken time reversal symmetry. In this regime, technically, the system is in the topological phase with non-zero Chern number and protected chiral edge modes. However, for larger $M$ this topological phase is protected by a gap that becomes exponentially small, thus limiting its potential usefulness.    

We may conclude that even though the $\cT$-breaking gap quickly becomes  small as the thickness of the flakes increases, the current-phase relation continues to display signatures of the spontaneous $\cT$ breaking in this limit. Intuitively, this behavior can be understood by noting that while the supercurrent is controlled by the properties of the interface, the spectral gap, which is a bulk property, is only weakly affected by the conditions at the surface.
We thus expect Fraunhofer interference patterns and fractional Shapiro steps discussed in the context of monolayer-thin flakes to remain a useful probe of spontaneous $\cT$ breaking in the limit of thicker flakes. Spectroscopies, on the other hand, will be most useful when applied in the limit of monolayer-thin flakes.   

\section{Discussion and conclusions}

Recent theoretical proposals suggest that twisted cuprate bilayers can form topological phases with superconducting critical temperatures that are close to the bulk critical temperature of the parent compounds. These exciting predictions beg the question: what would be an unequivocal signature of time reversal symmetry breaking -- and of a high-temperature topological superconducting phase? The full answer likely requires complementary evidence from different experimental quarters. Yet, quantum transport through the bilayer might be one of the simplest and most natural starting points. 

Because each layer is a standalone superconductor, the twisted structure forms a Josephson junction, provided that the interlayer coupling is weak. At the fundamental level, $\cT$ breaking becomes apparent in the current phase relation $J(\varphi)$ which reflects the double-minimum structure of the Josephson free energy $\cF(\varphi)$. But when the experimentally measurable critical current is analysed as a function of temperature and twist, somewhat surprisingly, we find the behavior to be smooth. In other words, a transition to the topological regime cannot be earmarked by simply looking at the critical current of an otherwise unperturbed junction. 

When the bilayer is subjected to certain external perturbations, however, the higher harmonic term corresponding to double Cooper pair tunneling can become manifest. Under the influence of an in-plane magnetic field, the junction shows a Fraunhofer oscillation pattern consistent with the new $\pi$-periodic term. Similarly, when the junction is driven with an electromagnetic radiation, fractional Shapiro steps appear in the current-voltage characteristic. 

It is to be noted that, strictly speaking, a detection of the $\pi$-periodic term is by itself not sufficient to identify the $\cT$-broken topological phase in our twisted system. An additional requirement is that the second harmonic term $J_{c2}$ comes with the opposite sign relative to the fundamental term $J_{c1}$, as indicated in Eq.\ \eqref{e4} and that  $2J_{c2}> J_{c1}$, which is the condition for the onset of the $\cT$-broken phase. Our theoretical considerations show that natural models of twisted bilayers produce the correct signs of $J_{c1,2}$, required to enter the $\cT$-broken phase. In addition, we find that clearly observable deviations from conventional Fraunhofer and Shapiro responses occur only when the condition $2J_{c2}> J_{c1}$ is well satisfied. This suggests that such effects can be taken as strong signatures of the topological phase.  

The current-phase relation describes supercurrent through the `bulk' of the 2D surface of the constituent twisted superconductors. In this sense, the Fraunhofer patterns and Shapiro steps studied in this work are a materialization of the bulk physics of the 2D topological phase. Chiral edge modes, mandated by the bulk boundary correspondence, are another feature of the topological phase. Direct examination of the edges through transport or spectroscopic measurements would provide useful complementary insights into the problem. From the view point of experiments, this is arguably a more difficult task. We defer a discussion of such aspects to a future work.

\section*{Acknowledgments}

We are grateful to Philip Kim, Jed Pixley, Andrew Potter, Pavel Volkov, Ziliang Ye, and Frank Zhao for illuminating discussions and correspondence. This research was supported in part by NSERC and the Canada First Research Excellence Fund, Quantum Materials and Future Technologies Program.


\bibliography{didJ}


\appendix

\section{Interlayer supercurrent analysis}\label{app1}

Some additional insights into the temperature dependence of the interlayer supercurrent discussed in Sec.\ II can be gained by analysing the Josephson relation \eqref{e5} in more detail. To this end it is useful to write down an explicit expression for the two positive eigenvalues of the BdG Hamiltomian \eqref{h4}:
\begin{equation}\label{a1}
E_{\bk\pm}=\sqrt{(\Delta_{\bk 1}^2+\Delta_{\bk 2}^2)/2 + \xi_\bk^2 + g^2 \pm D_\bk(\varphi)}
\end{equation}
where $D_\bk^2(\varphi)=(\Delta_{\bk 1}^2-\Delta_{\bk 2}^2)^2/4 +g^2(\Delta_{\bk 1}^2+\Delta_{\bk 2}^2+4\xi_\bk^2-2\Delta_{\bk 1}\Delta_{\bk 2}\cos{\varphi})$. Noting that the phase only enters through the cosine term in $D_\bk^2(\varphi)$ it is possible, with use of Eq.\ \eqref{h8}, to express the supercurrent as
\begin{equation}\label{a2}
J(\varphi)=-\sin{\varphi}{eg^2\over 2\hbar}\sum_\bk{\Delta_{\bk 1}\Delta_{\bk 2}\over D_\bk(\varphi)}\sum_{a=\pm} {a\over E_{\bk a}}\tanh{{1\over 2}\beta E_{\bk a}}.
\end{equation}
The superfluid stiffness follows from Eq.\ \eqref{e8},
\begin{equation}\label{a3}
\rho_s={eg^2\over 2\hbar}\sum_\bk{\Delta_{\bk 1}\Delta_{\bk 2}\over D_\bk(0)}\sum_{a=\pm} \left[{-a\over E_{\bk a}}\tanh{{1\over 2}\beta E_{\bk a}}\right]_{\varphi=0}.
\end{equation}
Noting that $E_{\bk +}>E_{\bk -}$ it is easy to see that the last term $\sum_a[\dots]$ in the above equation is positive for all $\beta$ which justifies the expression given in Eq.\ \eqref{s1} of the main text.

By linearizing the quasiparticle dispersion near the Dirac points -- performing the `nodal approximation' -- it is possible to extract from Eq.\ \eqref{a3} the functional from of the low-$T$ behavior, 
\begin{equation}\label{a4}
\rho_s(T)\simeq \rho_s(0) + a_\theta T-b_\theta T^3
\end{equation}
with non-negative $\theta$-dependent coefficients $a_\theta$ and $b_\theta$. This expression is valid for twist angles away from $\theta_M$ and indicates that for $a_\theta>0$ the $\rho_s(T)$ initially grows with temperature, as already observed Figs.\ \ref{fig4} and \ref{fig5}. The same calculation indicates a logarithmic contribution $\sim \ln{(k_B T/g)}$ when $\theta=\theta_M$. This is also evident in Fig.\ \ref{fig4}b.

The analysis of the critical current is more complicated because $J(\varphi)$ generally attains its maximum at some generic phase angle $\varphi_c$ which must be first determined. This in general can only be done numerically. Some analytical progress can be made by noting that at least for small twist angles $\varphi_c$ tends to be close to $\pi/2$. One may thus approximate $J_c\simeq J(\pi/2)$ and then estimate $J(\pi/2)$ using the nodal approximation as before. One finds exponentially activated behavior at the lowest temperatures reflecting the spectral gap that develops in the system at the large external phase bias. At somewhat higher temperatures $J(\pi/2)$
behaves as in Eq.\ \eqref{a4} but the log divergence near $\theta_M$ is now absent. This behavior is indeed observed in Fig.\ \ref{fig4}a.


\section{Expansion in powers of $g$}\label{app3}

We supply here some details leading to estimates quoted in Eqs.\ (\ref{s2},\ref{s3}). 

To estimate $J_c$ near zero twist angle we evaluate the current given in Eq.\ \eqref{a2} at $\theta=0$ to leading order in interlayer coupling $g$. In this regime the current is dominated by ordinary Josephson tunneling, which is a $g^2$ process, and we can thus set $g=0$ inside the momentum sum. At $T=0$ this gives
\begin{equation}\label{c1}
J(\varphi)=\sin{\varphi}{eg^2\over \hbar}\sum_\bk{\Delta^2\cos^2{2\alpha_\bk}\over 2(\xi_\bk^2+\Delta^2\cos^2{2\alpha_\bk})^{3/2}}.
\end{equation}
In this approximation the maximum current occurs at $\varphi=\pi/2$ and thus  $J_c= J(\pi/2)$. We replace the momentum sum by an integral as described in Appendix \ref{app2} below to obtain
\begin{equation}\label{c2}
J_c\simeq{eg^2\over \hbar}N_F\int_0^{2\pi} d\alpha\int_0^{x_c} dx{\cos^2{2\alpha}\over (x^2+\cos^2{2\alpha})^{3/2}}
\end{equation}
with $x_c={\epsilon_c/\Delta}$.
In the physically relevant limit $x_c\to\infty$ the integrals can be evaluated analytically which leads to Eq.\ \eqref{s2} with $C_2=1$.

The same procedure applied at twist angle $\theta=45^\circ$ gives vanishing current, consistent with the notion that single Cooper pair tunneling is disallowed in this limit. To obtain a meaningful estimate for $J_c$ one thus needs to carry the expansion to order $g^4$ which captures double-pair tunneling processes. Working once again at $T=0$ we find
\begin{equation}\label{c3}
J(\varphi)=\sin{\varphi}{eg^2\over \hbar}\sum_\bk{\Delta^2\sin{4\alpha_\bk}\over 2(E_{\bk+}+E_{\bk-})E_{\bk+}E_{\bk-}}.
\end{equation}
Expanding the denominator to order $g^2$, collecting all non-vanishing terms, and noting that in this case the current maximum occurs at $\varphi=\pi/4$ one arrives at Eq.\ \eqref{s3} with the constant given by
\begin{equation}\label{c2}
C_4=\int_0^{2\pi} {d\alpha\over 2\pi}\sin^2{4\alpha}\int_0^{x_c} dx{3(\epsilon_++\epsilon_-)^2-2-4x^2\over (\epsilon_++\epsilon_-)^3(\epsilon_+\epsilon_-)^3},
\end{equation}
where $\epsilon_\pm=\sqrt{2x^2+1\pm\cos{4\alpha}}$. Numerical integration in the limit $x_c\to\infty$ gives $C_4\simeq 0.55$.

%
\begin{figure}[t]
  \includegraphics[width=7.6cm]{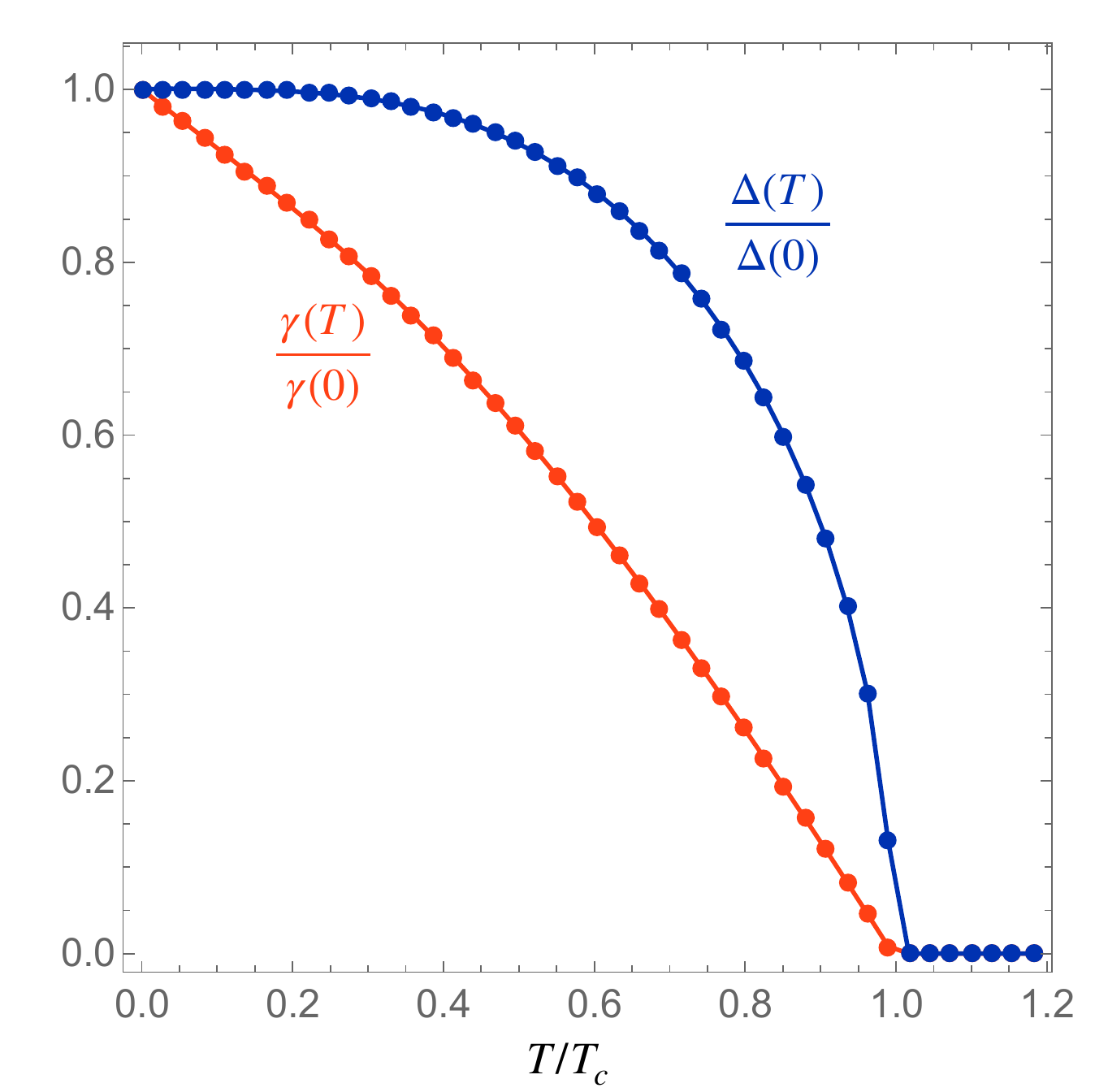}
  \caption{Temperature dependence of the superconducting gap amplitude (blue) calculated by numerically iterating the gap equation \eqref{h11}, and the in-plane superfluid stiffness $\gamma(T)$ (red) evaluated as the coefficient of the $q^2$ term in Eq.\ \eqref{a12}.}
  \label{fig9}
\end{figure} 

\section{In-plane superfluid stiffness}\label{app2}

To obtain the $\gamma$ coefficient in the single-layer free energy Eq.\ \eqref{f2} we begin from the relevant real-space BdG Hamiltonian, 
\begin{equation}\label{a5}
h=\begin{pmatrix}
  h_0 & \Delta(\br) \\
 \Delta(\br)^* & -h_0  
\end{pmatrix}
\end{equation}
where $h_0=-\hbar^2\nabla^2/2m -\mu$ and $\Delta(\br)=e^{2iqx}\hat\Delta_d$. The exponential factor in the order parameter sets up a uniform superflow along the $x$ direction with the amplitude proportional to $q$. The coefficient $\gamma$ measures the free energy cost of this supercurrent per unit volume. To proceed it is useful to pass to a new gauge, taking $c_{\uparrow\br}\to e^{-iqx}c_{\uparrow\br}$ and $c^\dag_{\downarrow\br}\to e^{iqx}c^\dag_{\downarrow\br}$. This removes the phase factor from $\Delta$ and transforms the kinetic term as $\pm h_0\to\mp(\hbar^2/2m)(\nabla\mp i\bq)^2$
where $\bq=(q,0,0)$. The advantage of this gauge is that the Hamiltonian becomes translation invariant and can hence be expressed in the momentum space as 
\begin{equation}\label{a6}
h_\bk=\begin{pmatrix}
  \xi_{\bk-\bq} & \Delta_\bk \\
 \Delta_\bk^* & -\xi_{\bk+\bq}
\end{pmatrix},
\end{equation}
where $\xi_\bk=\hbar^2\bk^2/2m -\mu$ and $\Delta_\bk=\Delta_0\cos(2\alpha_\bk)$. The free energy can now be calculated from the energy eigenvalues 
\begin{equation}\label{a7}
E_{\bk\pm}={\xi_{\bk+\bq}-\xi_{\bk-\bq}\over 2} \pm\sqrt{\left({\xi_{\bk+\bq}+\xi_{\bk-\bq}\over 2}\right)^2+\Delta_\bk^2}
\end{equation}
and reads
\begin{equation}\label{a8}
F=E_0-{1\over \beta}\sum_{\bk,a=\pm}\ln[2\cosh(\beta E_{\bk a}/2)],
\end{equation}
where $E_0$ is independent of $\bq$.

The quantity of interest is the in-plane phase stiffness which follows from the $\gamma$ term in Eq.\ \eqref{a5} upon taking $\psi(\br)=\psi_0 e^{i\varphi(\br)}$. For the uniform supercurrent defined by $\varphi(\br)=2qx$ this takes the form
$\gamma|\nabla\psi|^2=4\gamma\psi^2_0 q^2$. Therefore, for a weak supercurrent, we seek the coefficient of $q^2$ in the expansion of free energy \eqref{a8} in powers of $q$. Before we proceed with the expansion it is useful to switch perspective slightly and view Eqs.\ (\ref{a6}-\ref{a8}) as describing a lattice model; specifically we shall henceforth regard $\xi_\bk$ and $\Delta_\bk$ as lattice dispersion and gap function, respectively, defined in the Brillouin zone appropriate for the square CuO$_2$ lattice. This point of view avoids difficulties down the road where the continuum model would show various ultraviolet-divergent $k$-space integrals. By contrast no such divergences appear in the lattice model because the BZ provides a natural UV cutoff. 

Expanding the free energy \eqref{a8} to second order in $q$ we find
\begin{widetext}
\begin{equation}\label{a9}
F\simeq F_0-{1\over 2} q^2\sum_{\bk}\left[\left({\partial^2\xi_\bk\over\partial k_x^2}\right)\left(1-{\xi_\bk\over \epsilon_\bk}\tanh{\beta \epsilon_{\bk}\over 2}\right) - {\beta\over 2}\left({\partial\xi_\bk\over\partial k_x}\right)^2{\rm sech}^2{\beta \epsilon_{\bk}\over 2}
\right],
\end{equation}
where $\epsilon_\bk=\sqrt{\xi_\bk^2+\Delta_\bk^2}$ is the quasiparticle excitation energy. 
For a $d$-wave SC it is useful to follow Ref.\ \cite{Sheehy2004} and integrate the first term by parts. After some algebra we obtain
\begin{equation}\label{a10}
F\simeq F_0+{1\over 2} q^2\sum_{\bk}\left({\partial\xi_\bk\over\partial k_x}\right)
\left[{\Delta_\bk^2\over\epsilon_\bk^2}\left({\partial\xi_\bk\over\partial k_x}\right)-
{\Delta_\bk\xi_\bk\over\epsilon_\bk^2}\left({\partial\Delta_\bk\over\partial k_x}\right)\right]
\left[{1\over\epsilon_\bk}\tanh{\beta \epsilon_{\bk}\over 2} - {\beta\over 2}{\rm sech}^2{\beta \epsilon_{\bk}\over 2}
\right].
\end{equation}
This expression, while seemingly more complicated than Eq.\ \eqref{a9}, has several desirable features in terms of numerical evaluation. First, the coefficient of $q^2$ explicitly vanishes when $\Delta_\bk=0$; indeed one expects superfluid stiffness to be zero in the normal metal limit. Second, because of the powers of the excitation energy $\epsilon_\bk$ present in various denominators, it is clear that the largest contributions to the momentum sum comes from the nodal regions where $\epsilon_\bk\to 0$. These properties facilitate a straightforward evaluation of the $k$-space sum and we can now safely revert back to the continuum approximation. To this end we consider free energy density $f=F/V$, where $V$ is the sample volume, and convert the $k$-space sum to an integral using the standard prescription 
\begin{equation}\label{a11}
{1\over V}\sum_\bk \,\to\, {1\over (2\pi)^2}\int d^2k \, \to \,
{1\over (2\pi)^{2}}\int_0^{2\pi}d\alpha\int kdk
\, \to \, N_F\int_0^{2\pi}d\alpha\int_{-\epsilon_c}^{\epsilon_c}d\xi.
\end{equation}
Here $N_F={m/(2\pi\hbar)^{2}}$ is the density of states at the Fermi level and in the last step we restricted the integration to within a cutoff $\epsilon_c$ around the Fermi level which is now a legitimate approximation because the integrand is strongly peaked here. With these simplifications the free energy density expansion can be written as 
\begin{equation}\label{a12}
f\simeq f_0+{q^2\over \pi^2}\int_0^\pi d\alpha(\cos{\alpha}\cos{2\alpha})^2
\int_0^{\epsilon_c}d\xi{\Delta^2\mu\over\epsilon^3}\left[\tanh{\beta\epsilon\over 2}-{\beta\epsilon\over 2}{\rm sech}^2{\beta\epsilon\over 2}\right],
\end{equation}
\end{widetext}
where $\epsilon=\sqrt{\xi^2+\Delta^2\cos^2{2\alpha}}$. The integrals indicated in Eq.\ \eqref{a12} lend themselves to a straightforward numerical evaluation and lead to the well-known curves for in-plane superfluid stiffness of a $d$-wave SC with the characteristic $T$-linear dependence at low temperatures. An example of this behavior is given in Fig.\ \ref{fig9} along with the temperature dependence of the gap function $\Delta(T)$, which is required as an input for this calculation.


\end{document}